\documentstyle[amssymb,epsfig,12pt]{article}
\begin{document}

\title{A process-reconstruction analysis of market fluctuations}
\author{R. Vilela Mendes\thanks{%
Laborat\'{o}rio de Mecatr\'{o}nica, DEEC, Instituto Superior T\'{e}cnico,
Av. Rovisco Pais, 1096 Lisboa Codex, Portugal (vilela@cii.fc.ul.pt)} \thanks{%
Zentrum f\"{u}r interdisziplin\"{a}re Forschung, Universit\"{a}t Bielefeld,
Wellenberg 1, 33615 Bielefeld, Germany}, R. Lima\thanks{%
Centre de Physique Th\'{e}orique CNRS, Luminy, Case 907, F13288 Marseille
Cedex 9, France (lima@cpt.univ-mrs.fr)} \footnotemark[2]  and T. Ara\'{u}jo%
\thanks{%
Dept. Economia, ISEG, R. Miguel Lupi 20, 1200 Lisboa, Portugal
(tanya@iseg.utl.pt)} \footnotemark[2] }
\date{}
\maketitle

\begin{abstract}
The statistical properties of a stochastic process may be described (1)by
the expectation values of the observables, (2)by the probability
distribution functions or (3)by probability measures on path space. Here an
analysis of level (3) is carried out for market fluctuation processes. Gibbs
measures and chains with complete connections are considered. Some other
topics are also discussed, in particular the asymptotic stationarity of the
processes and the behavior of statistical indicators of level (1) and (2).
We end up with some remarks concerning the nature of the market
fluctuation process.
\end{abstract}

{\bf Keywords}: Market fluctuations, Gibbs measures, Chains with complete
connections

\section{Introduction}

When a physical phenomenon is measured with a set of instruments, what we
register is a sequence of values of some variable $X$

\[
\cdots X_{-2}X_{-1}X_{0}X_{1}X_{2}\cdots 
\]
which takes values in a space $Y$. We will call $Y$ the {\it state space}
and the space of sequences $Y^{{\Bbb Z}}$ the {\it path space}. Statistical
properties of the phenomenon may be described at three different levels:

(1) By the expectation values of the observables;

(2) By the probability measures on the state space $Y$;

(3) By the probability measures on path space $Y^{Z}$.

One obtains three different characterizations of the phenomenon which
represent successively finer levels of description of the statistical
properties. Borrowing a terminology used in large deviation theory\cite
{Stroock} \cite{Ellis}, we will call these three types of description,
respectively, {\it level 1, 2 and 3- statistical indicators.}

To obtain expectation values and probability measures we would require
infinite samples and a law of large numbers. For any finite sample we obtain
finite versions of the expectation values, the probability on state space
and the probability on path space which are called the {\it mean partial
sums, }the {\it empirical measures }(or empirical probability distribution
functions - pdf's) and the measures on the {\it empirical process}.

Level-1 and level-2 analysis are the most common ones and their statistical
indicators the most commonly quoted when a stochastic process is analyzed.
However to the same expectation values for the observables or to the same
pdf's, different processes may be associated. Therefore full understanding
of the process requires the determination of the level-3 indicators. Recent
advances have been obtained on the identification of processes, especially
in connection with the analysis of hydrodynamic turbulence data\cite{Ugalde} 
\cite{Lima1} \cite{Lima2} \cite{Collet} \cite{Chazottes}. In particular it
has been clarified that analysis and reconstruction of the process involves
two different but related steps. One is the identification of the {\it %
grammar} of the process, that is, the allowed transitions in the state space
or the subspace in path space that corresponds to actual orbits of the
system. The second step is the identification of the {\it measure}, which
concerns the occurrence frequency of each orbit in typical samples. Although
largely independent from each other, this two features have a related effect
on the constraints they impose on the statistical indicators.

Identification of grammars and measures (in particular Gibbs measures) has
been dealt with recently, in particular in the context of hydrodynamic
turbulence and other dynamical systems. Market fluctuations is an
interesting stochastic process\cite{Bouchaudb} \cite{Stanleyb}. 
Some analogies have been found between this
process and some of the features of turbulence data\cite{Breymann1} \cite
{Breymann2}. However, when statistical indicators are computed, it turns out
that the two processes are different\cite{Stanley2} \cite{Arneodo} \cite
{Stanley1}. Nevertheless the statistical tools that have been developed for
turbulence are mathematical devices which are not process-dependent and they
may be applied to any stochastic process process. Of course, underlying this
approach is the working hypothesis that statistical methods, by themselves,
are an appropriate tool to describe and reconstruct the market fluctuation
process. This hypothesis underlies the modern view of the {\it efficient
market}, namely the idea that the market appears to overreact in some
circumstances and underreact in others is pure chance \cite{Fama}. In other
words, the expected value of abnormal returns is zero. Contrariwise, if a
well defined deterministic pattern of over- and underreaction is ever found
then, in addition to chance, a behavioral component \cite{Barberis} \cite
{DeBondt} must always be included in any description of the market.
Behavioral trends, however, may turn out not inconsistent with a pure
statistical description if the different reaction times of the diverse
market components are taken into account, as well as the secondary reactions
of the components to each other moves \cite{Olsen}.

The emphasis on this paper will be on level-3 analysis and on the
reconstruction of the processes. Nevertheless we have also dedicated some
time to the computation, for market fluctuations, of the level-1 and level-2
statistical indicators used in the past for turbulence data. In particular
the behavior of some of these indicators already provides information on the
nature of the grammars. This analysis is carried out in Sect. 3. Sect. 4 is
dedicated to the search for a Gibbs measure and, once the long-memory
features of the market processes are exhibited, Sect. 5 attempts to describe
the processes in the framework of chains with complete connections.

However, the first step in the analysis of any stochastic process is to
inquire about the stationarity of the process and whether typical samples
are available. This is the subject of the next section.

\section{Is there an asymptotically stationary market fluctuation phenomenon?
}

Large samples of high-frequency finance data are now available. However
high-frequency data may not be the more appropriate data to begin
understanding the stochastic process that underlies the market mechanism.
This is because, when comparing minute to monthly variations for example,
one is comparing systems with very different compositions, trading agents
operating on the minute scale being in general different from those
operating in longer time scales. This is evidenced, for example, by the
different scaling laws for low and high-frequency data. In market data one
faces a complexity versus statistics trade-off. The high frequency data
certainly provides better statistics but it also involves the interplay of
many more reaction time scales and market compositions in the trading
process. For this reason, to ``purify'' as much as possible our samples, we
have decided to concentrate on daily data. The price to be paid for this
choice is the fact that, as compared for example with a large scale
hydrodynamics experiment, the available amount of one-day market fluctuation
data is relatively small. If, in addition, the data is non-stationary, the
chances to obtain a reliable statistical analysis would be rather slim.

Reliable application of statistical mechanics tools to any kind of signal,
presupposes that two conditions are fulfilled. First, that the process that
generates the data has some kind of underlying stationarity or asymptotic
stationarity. Second, that the time sequence that is presented to the
analysis is a typical sample of the process. The second condition, of
course, we can only hope that it is realized and to improve our belief in
this condition several different signals of a similar nature should be
analyzed (several different stocks, or currencies or markets). As to the
first condition it requires some preprocessing of the data. We will
concentrate in this paper in the daily fluctuation data of industrial stocks
and indexes and the objective is to try to extract the features of the
market process that acts on them. We look at each stock as an experimental
probe that, while reacting to the market pressures, may reveal some of the
mechanisms of the market process.

Market prices are by nature non-stationary entities. They fluctuate, they
have general trends that depend on the general state of the economy, on the
total amount of capital flowing to the market, on the general acceleration
of the economy, on long and medium term political decisions and
expectations, etc. Nevertheless, our hypothesis is that, if all these global
factors are extracted from the data, there are still some invariant features
that characterize this peculiar human phenomenon.

\begin{figure}[htb]
\begin{center}
\psfig{figure=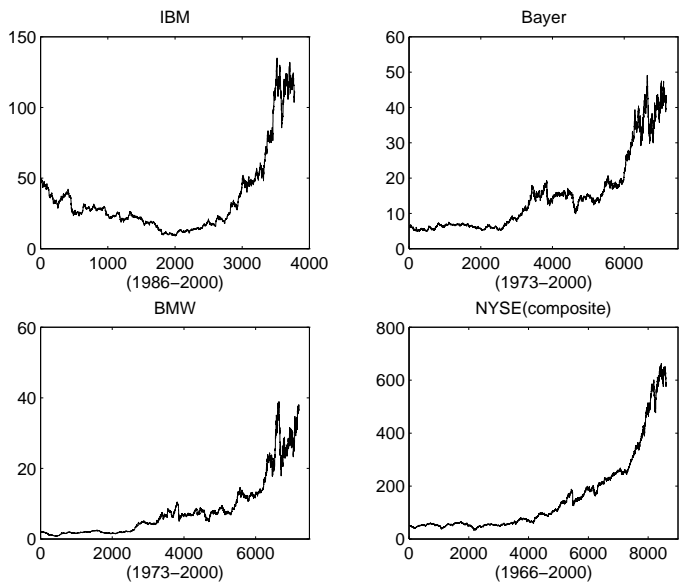,width=9truecm}
\end{center}
\caption{Daily price fluctuation data}
\end{figure}

The type of data that will be analyzed is displayed in Fig.1 that shows
daily price data $p(t)$ for three stocks and the NYSE composite index. Its
non-stationary nature is very apparent. The first step is to extract the
general trend. This is done, in a smooth way by a polynomial fit $q(t)$
(Fig.2 shows an example, where a 7-degree polynomial is used). Fig.3 shows
the difference $p(t)-q(t)$. Clearly the data is still very far from
stationary, because due to the market volume acceleration recent
fluctuations carry a much larger weight. Therefore the last step is a
rescaling of the data, by the average $<p(t)>$, that is 
\begin{equation}
x(t)=\left( p(t)-q(t)\right) \frac{<p(t)>}{q(t)}  \label{2.1}
\end{equation}
are the signals to be analyzed.. They are shown in Fig.4. To anyone used to
examine turbulence data, it looks as if the market signals are now somewhat
stable. That does not mean, of course, that they are stationary in the
strict sense. However it suggests that in spite of currency adjustments,
increased number of players, trade volumes and other macroeconomic
indicators, there is something more or less permanent in this human game.

\begin{figure}[htb]
\begin{center}
\psfig{figure=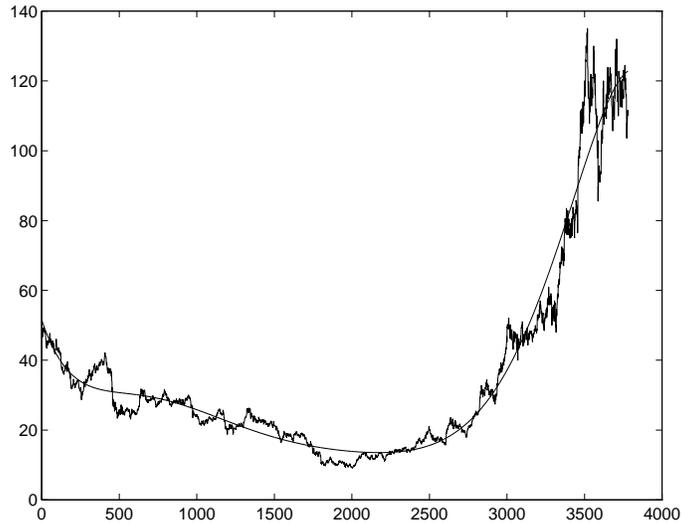,width=9truecm}
\end{center}
\caption{Detrending by a polynomial}
\end{figure}

\begin{figure}[htb]
\begin{center}
\psfig{figure=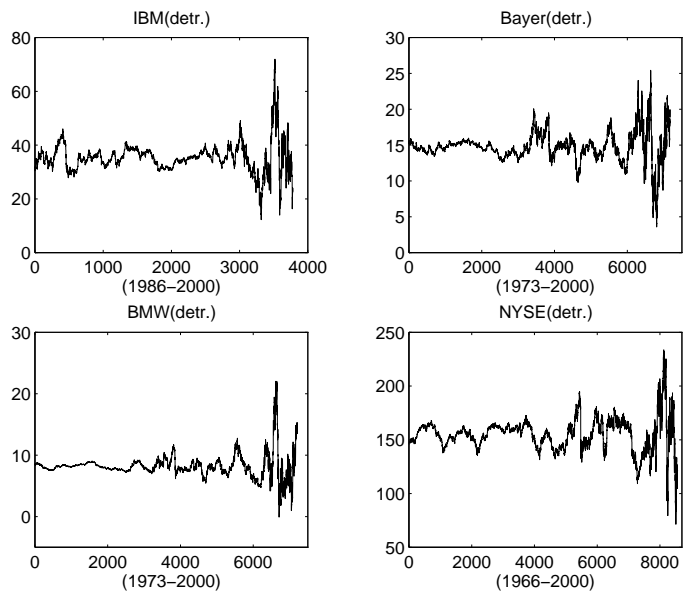,width=9truecm}
\end{center}
\caption{Detrended data}
\end{figure}

\begin{figure}[htb]
\begin{center}
\psfig{figure=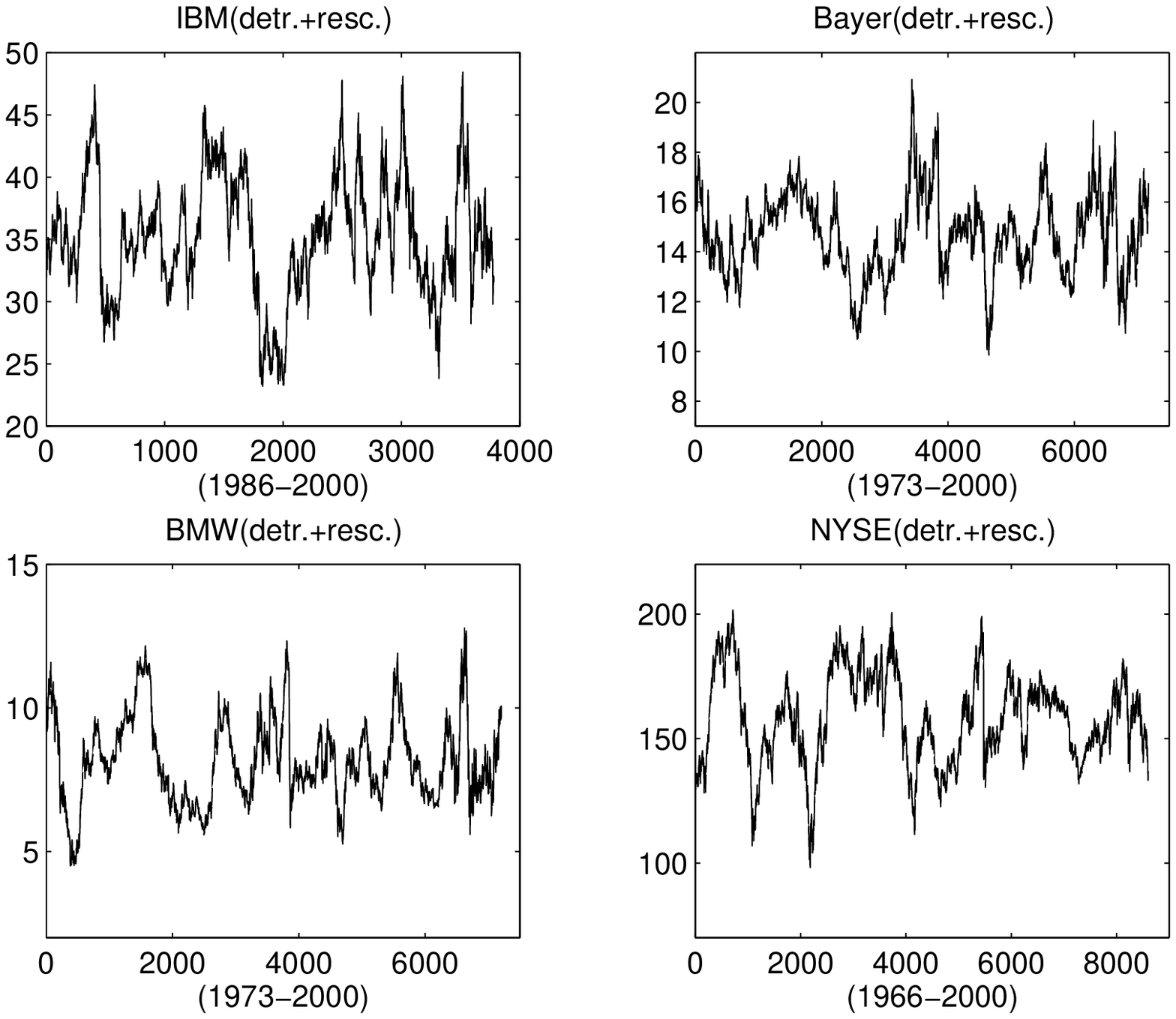,width=9truecm}
\end{center}
\caption{Detrended and rescaled data}
\end{figure}

Detrending and rescaling of the data is important because we will be
analyzing price differences over large time intervals. For one-day
differences of log-price, the results would be identical to those obtained
from the raw data. Detrending and rescaling the data, the overall amplitude
of price fluctuations becomes reasonably uniform over the time span of the
data. However the process is not (locally) stationary, as seen in Figs.5 and
6 that show the strong variation in time of the volatility (here defined as
the standard deviation of the price fluctuations). The two figures on the
left show the standard deviation computed on a sliding time window of 10
days. On the right one compares the cumulative standard deviation for the
rescaled (full line) and the non-rescaled data (dashed line). It is quite
apparent that only the rescaled data has the chance to belong to an
asymptotically stationary process. Once the data is detrended and rescaled
there is in fact no evidence\cite{Laster} for an abnormal increase, in
recent times, of the volatility in the underlying process.

\begin{figure}[htb]
\begin{center}
\psfig{figure=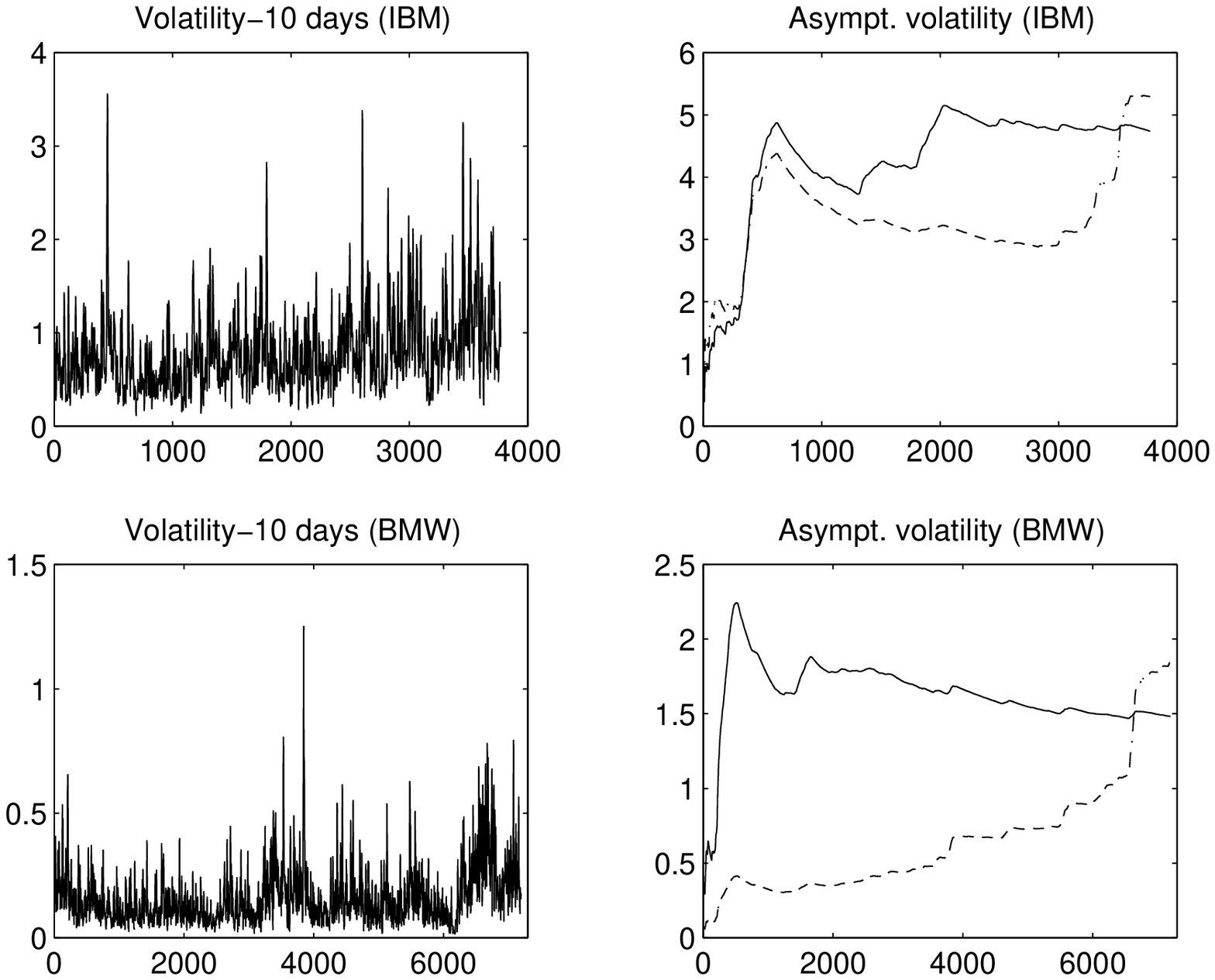,width=9truecm}
\end{center}
\caption{Ten-days window volatility and comparison of asymptotic volatility
for the rescaled and non-rescaled data (IBM and BMW)}
\end{figure}

\begin{figure}[htb]
\begin{center}
\psfig{figure=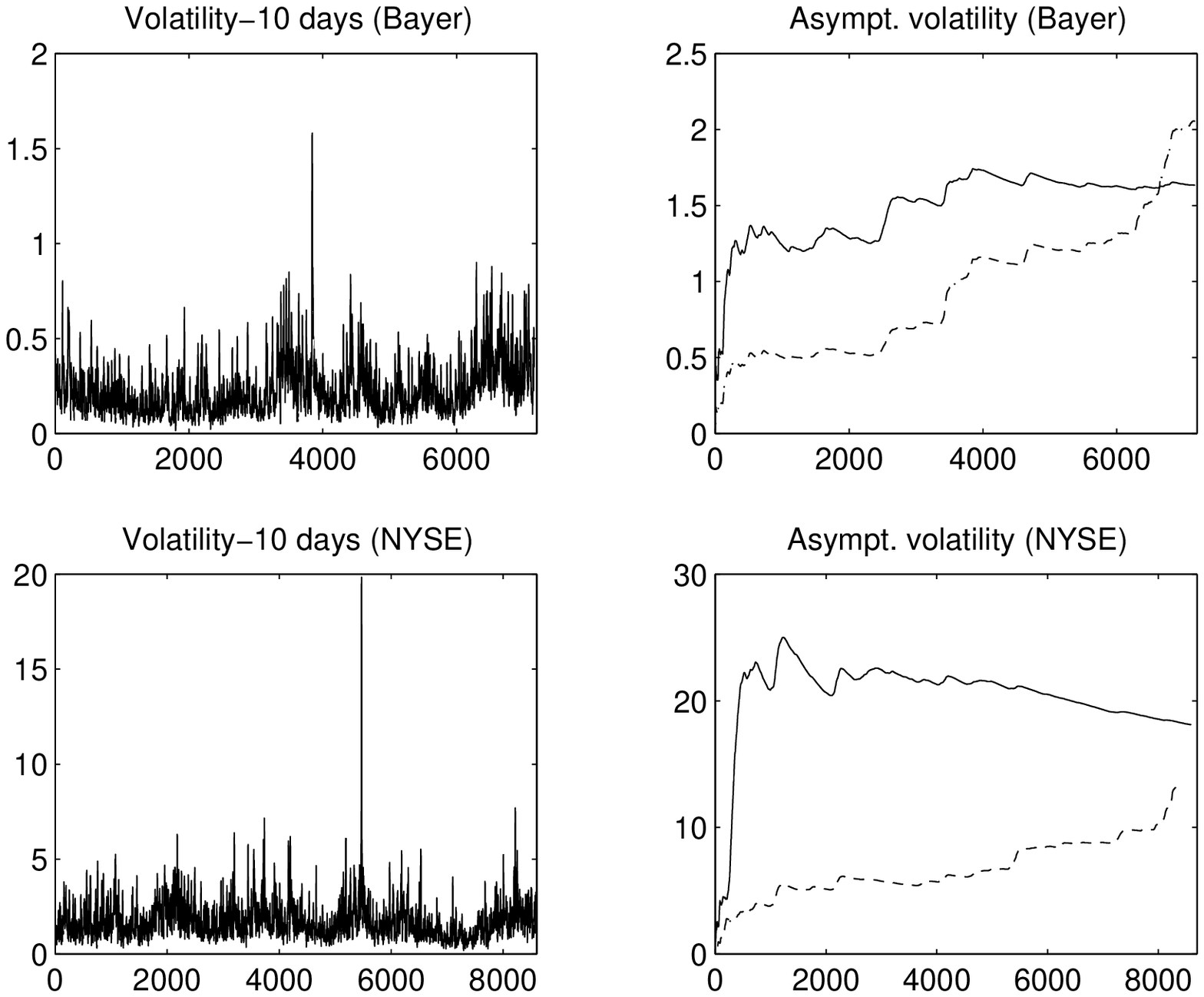,width=9truecm}
\end{center}
\caption{Same as in Fig.5 for Bayer and NYSE}
\end{figure}

A direct test of stationarity of the detrended and rescaled data was
obtained by coding with a 5-symbols alphabet (as explained in Sect. 4).
Then, computing the entropies of multi-symbol words, in the first and the
second half of the samples, no significant difference is found.

\section{Statistical indicators for typical samples}

Here we concentrate on level-1 and level-2 analysis of the regularized
samples discussed in Sect.2, that is, we compute quantities related to
averages values and to probability distribution functions (pdf 's). The
level-3 analysis of the processes will be done in the latter sections.

The main variables that are used to construct the statistical indicators are
the differences of log-prices 
\begin{equation}
r(t,n)=\log p\left( t+n\right) -\log p\left( t\right)  \label{3.1}
\end{equation}
sometimes called the $n-$days return. For each experimental sample, three
main statistical indicators are computed:

(i) The maximum (over $t$) of $r(t,n)$%
\begin{equation}
\delta \left( n\right) =\max_{t}\left\{ r(t,n)\right\}  \label{3.2}
\end{equation}

(ii) The moments of the distribution of $\left| r(t,n)\right| $%
\begin{equation}
S_{q}(n)=\left\langle \left| r(t,n)\right| ^{q}\right\rangle  \label{3.3}
\end{equation}
with $\left\langle \quad \right\rangle $ meaning the sample average

(iii) If inside a certain range, the moments satisfy 
\begin{equation}
S_{q}(n)\thicksim n^{\chi (q)}  \label{3.4}
\end{equation}
then the scaling exponent $\chi (q)$ is another important statistical
indicator.

\begin{figure}[htb]
\begin{center}
\psfig{figure=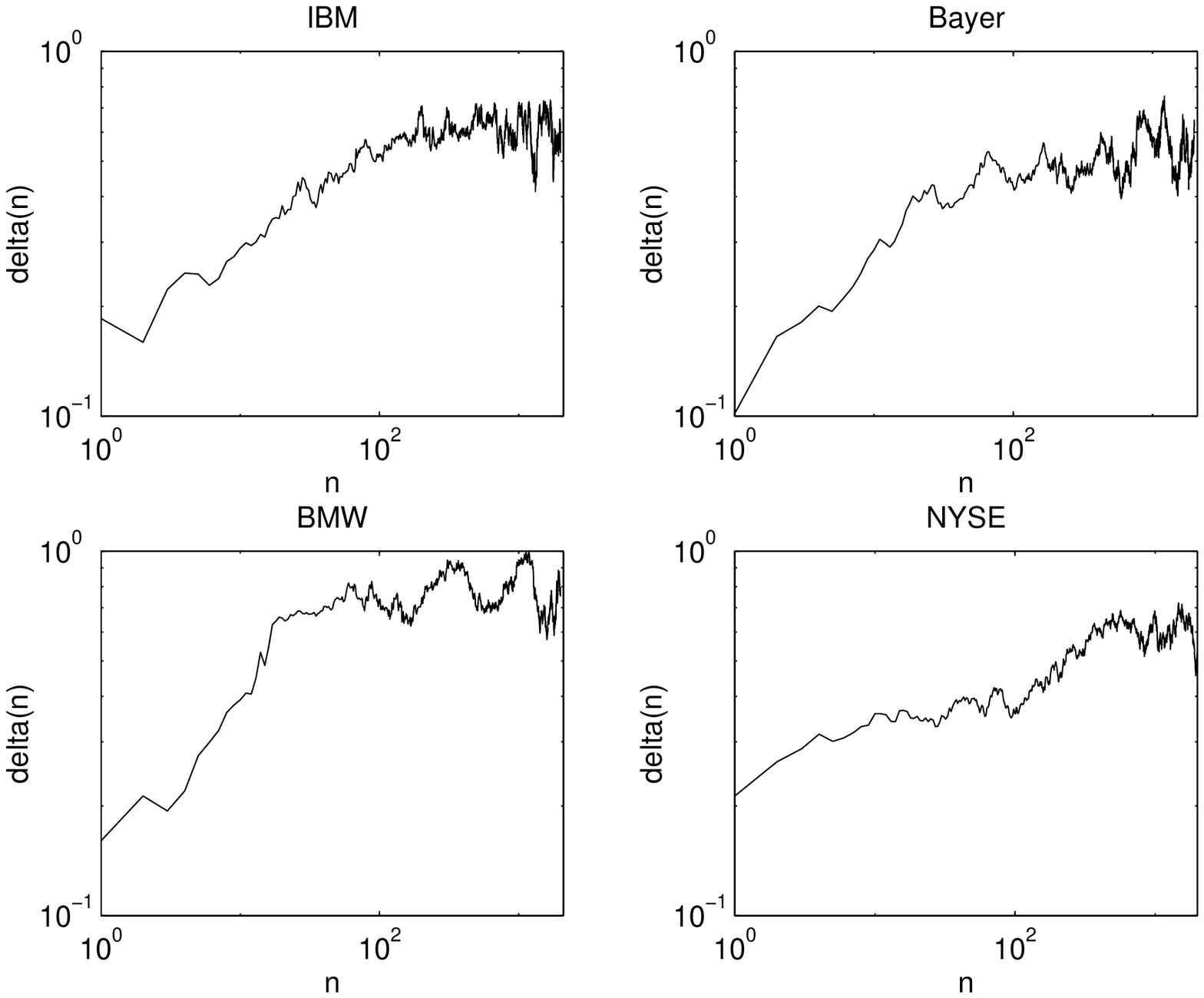,width=9truecm}
\end{center}
\caption{Maximum $\delta \left( n\right) $ of log-prices differences}
\end{figure}

The results obtained from our detrended and rescaled samples are displayed
in Figs.7 to 9. Fig.7 refers to $\delta \left( n\right) $ and Fig.8 shows $%
S_{q}(n)$ as a function of $n$ for different values of $q$ (from top to
bottom $q=1$ to $q=8$). The large fluctuations in $\delta \left( n\right) $
for large values of $n$ and in $S_{q}(n)$ for large $q$ are quite natural
given the size of the data samples.

\begin{figure}[htb]
\begin{center}
\psfig{figure=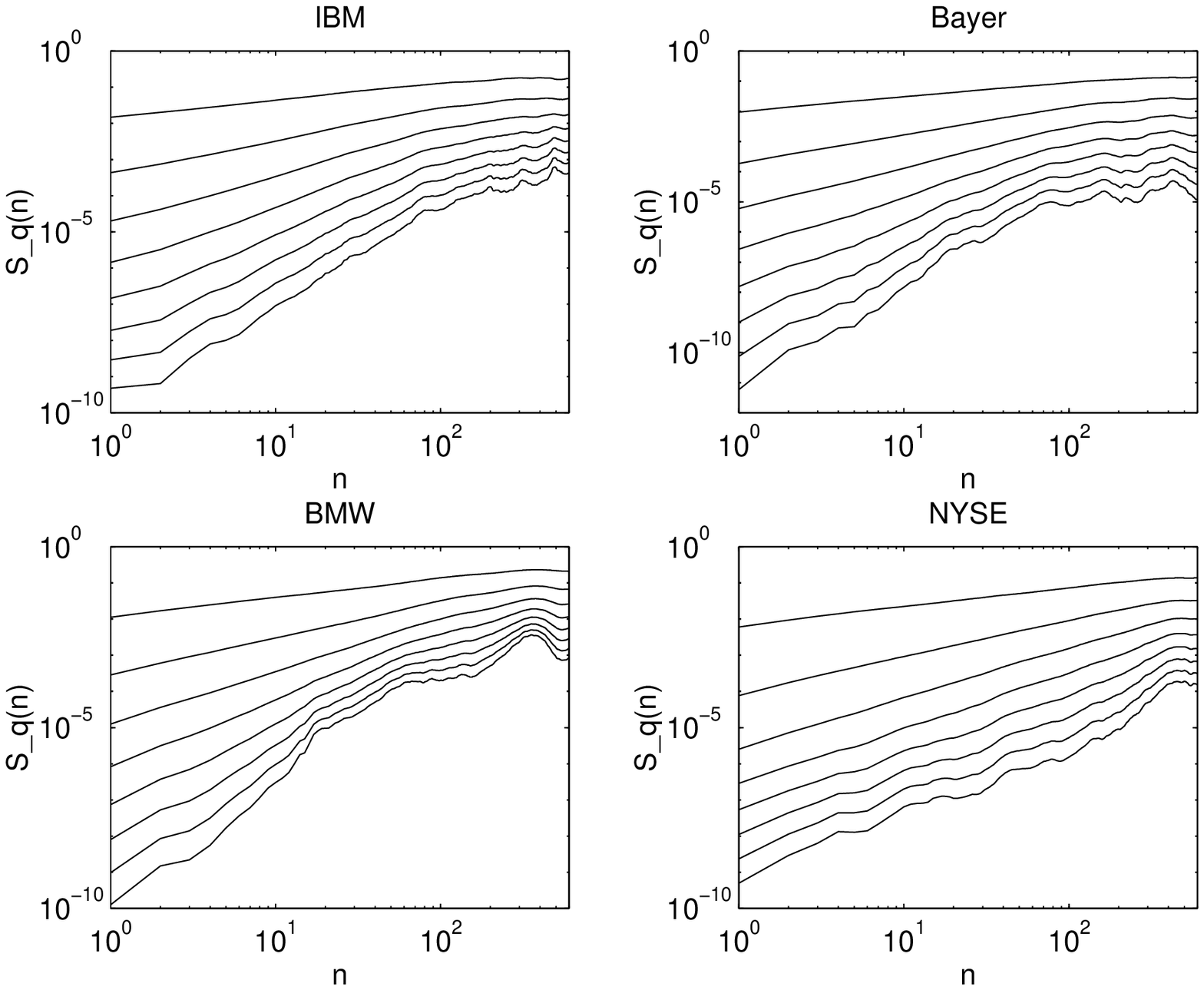,width=9truecm}
\end{center}
\caption{Moments of the $\left| r(t,n)\right| $ distribution}
\end{figure}

\begin{figure}[htb]
\begin{center}
\psfig{figure=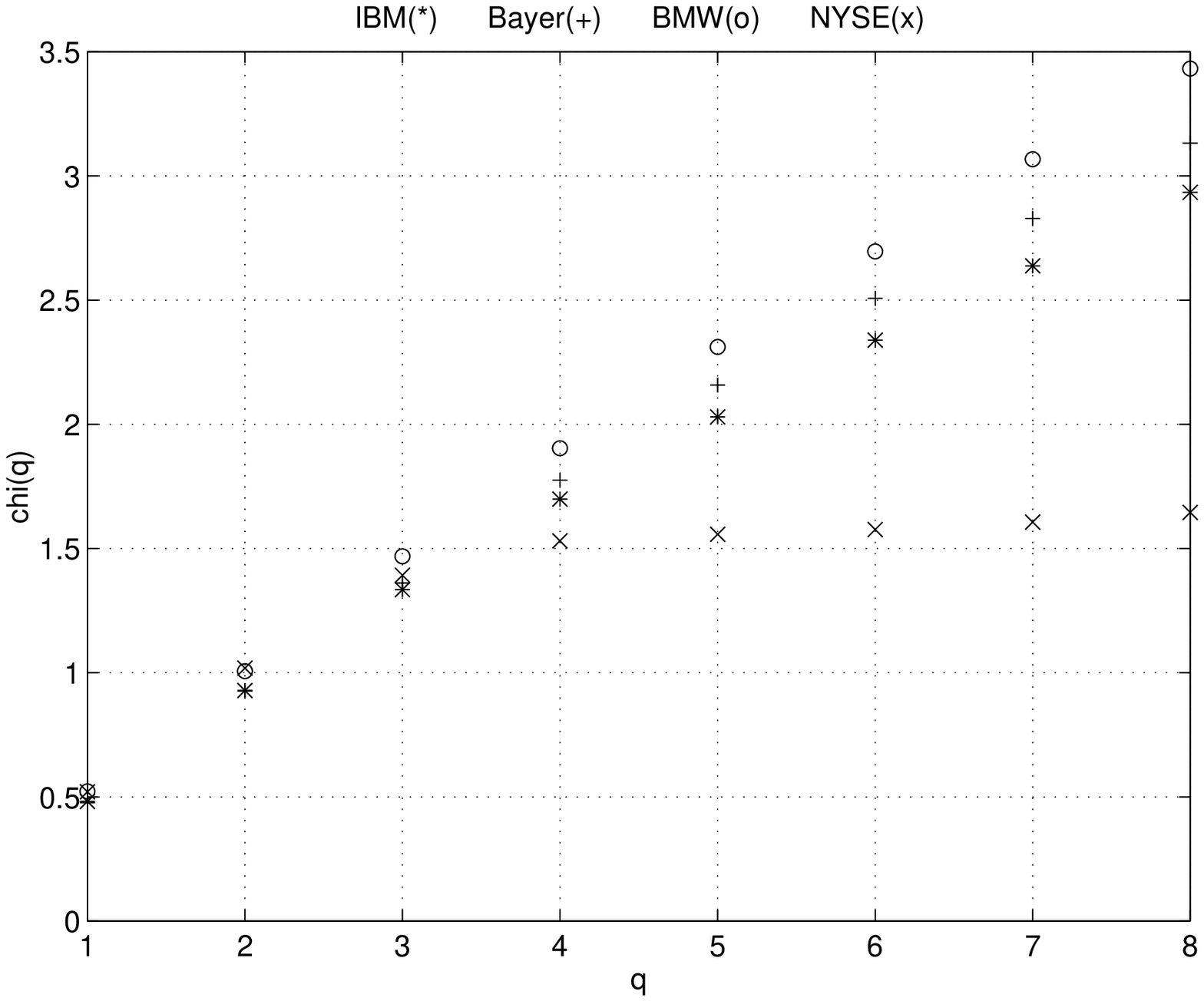,width=9truecm}
\end{center}
\caption{Scaling exponent $\chi (q)$}
\end{figure}

In the range $n=2$ to $n=60$ the moments follow an approximate power law of
the type of Eq.(\ref{3.4}) and from the behavior in this region we have
extracted the scaling exponent $\chi (q)$ shown in Fig.9. The main
conclusions from this analysis of the statistical indicators are:

(a) $\delta \left( n\right) $ is log-concave, that is, $\log \delta \left(
n\right) $ is concave as a function of $\log n$, increasing and probably
(with better statistics) asymptotically constant for large $r$;

(b) $S_{q}(n)$ is also an increasing log-concave function of $n$, allowing a
power law approximation in a limited range;

(c) The scaling law $\chi (q)$ is an increasing concave function of $q$;

(d) For all samples, $\chi (1)$ computed in the scaling region ($n=2$ to $%
n=60$) is very close to 0.5;

(e) The scaling properties of the NYSE index seem somewhat different from
those of the other stocks. However this is only apparent for $q\geq 5$,
where poor statistics effects may already be felt.

From this analysis one also obtains precise statements concerning the
similarities and differences between hydrodynamic turbulence and the market
fluctuation process. Properties (a) to (c) are shared by the turbulence
data, although the numerical values of the statistical indicators are quite
different. For example, for turbulence data $\chi (1)=\frac{1}{3}$ whereas
here $\chi (1)\thickapprox 0.5$, showing the essentially uncorrelated nature
of the signal for $n\geq 2$. The correlation function of one-day returns and
its absolute value 
\begin{equation}
C\left( r(1),T\right) =\left\langle r\left( t+T,1\right) r\left( t,1\right)
\right\rangle  \label{3.5}
\end{equation}
and 
\begin{equation}
C\left( \left| r(1)\right| ,T\right) =\left\langle \left| r\left(
t+T,1\right) \right| \left| r\left( t,1\right) \right| \right\rangle
\label{3.6}
\end{equation}
are shown in Fig.10. One sees that for $T\geq 2$ the returns are
uncorrelated, their correlation function remaining at the noise level. In
contrast the correlation for the absolute value remains non-negligible for a
longer time (at least up to $T=10$). This means that although the returns
are linearly uncorrelated, non-linear functions of the returns remain
correlated for longer periods.

\begin{figure}[htb]
\begin{center}
\psfig{figure=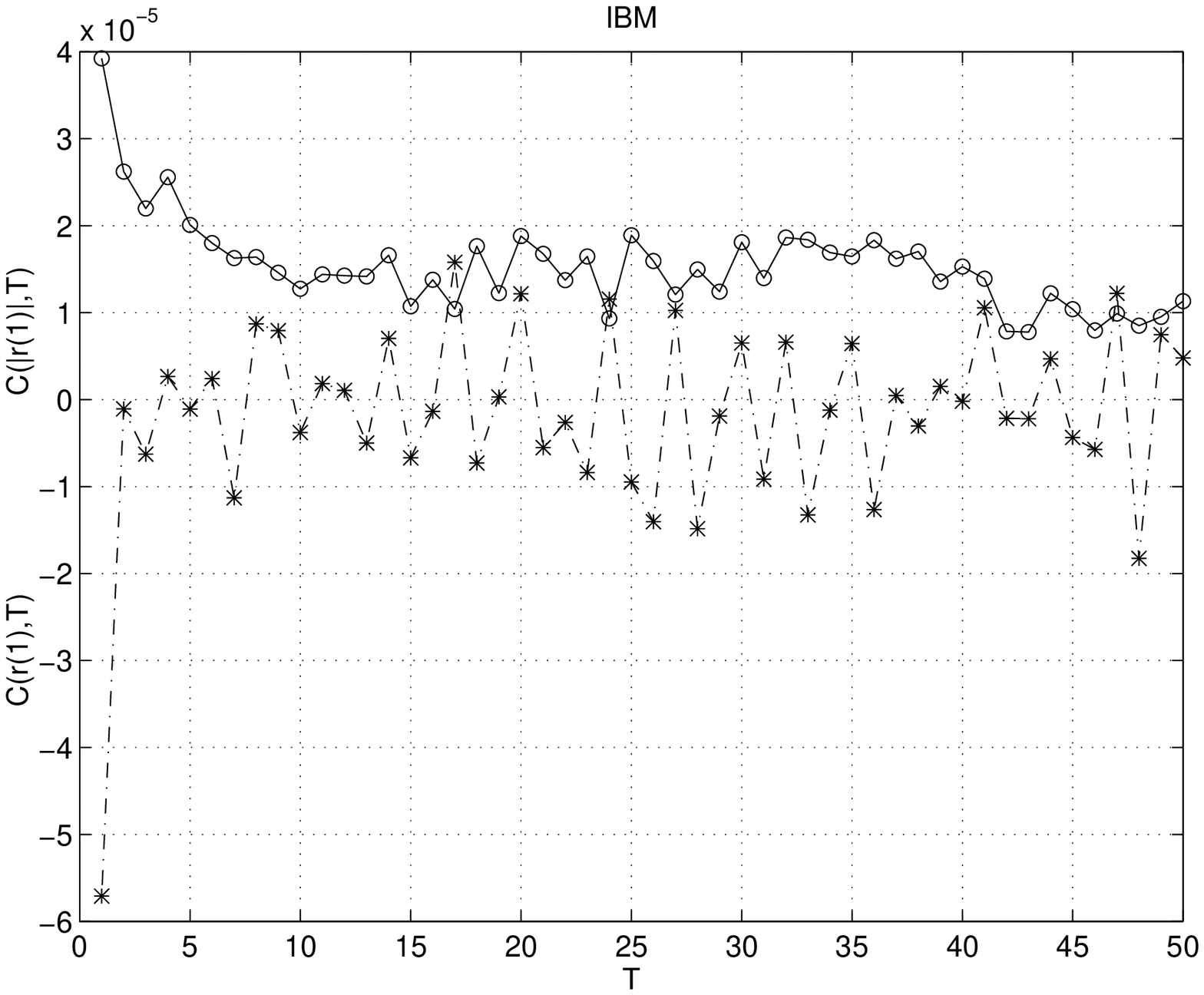,width=9truecm}
\end{center}
\caption{Correlation functions of $r(t,1)$ and $\left| r(t,1)\right| $}
\end{figure}

The behavior of the statistical indicators $\delta \left( n\right) $, $%
S_{q}(n)$ and $\chi (q)$ already has some strong implications on the level-3
features of the process, namely on the structure of its grammar. In fact,
without restrictions on the allowed transitions $\delta \left( n\right) $
and $S_{q}(n)$ would be independent of $n$ and $\chi (q)=0$ for all $q$. In
particular, property (a) implies that if the process is a topological Markov
chain the transitions allowed by the transition matrix $T$ must lie inside a
strictly convex domain around the diagonal of $T$\cite{Lima2}.

\begin{figure}[htb]
\begin{center}
\psfig{figure=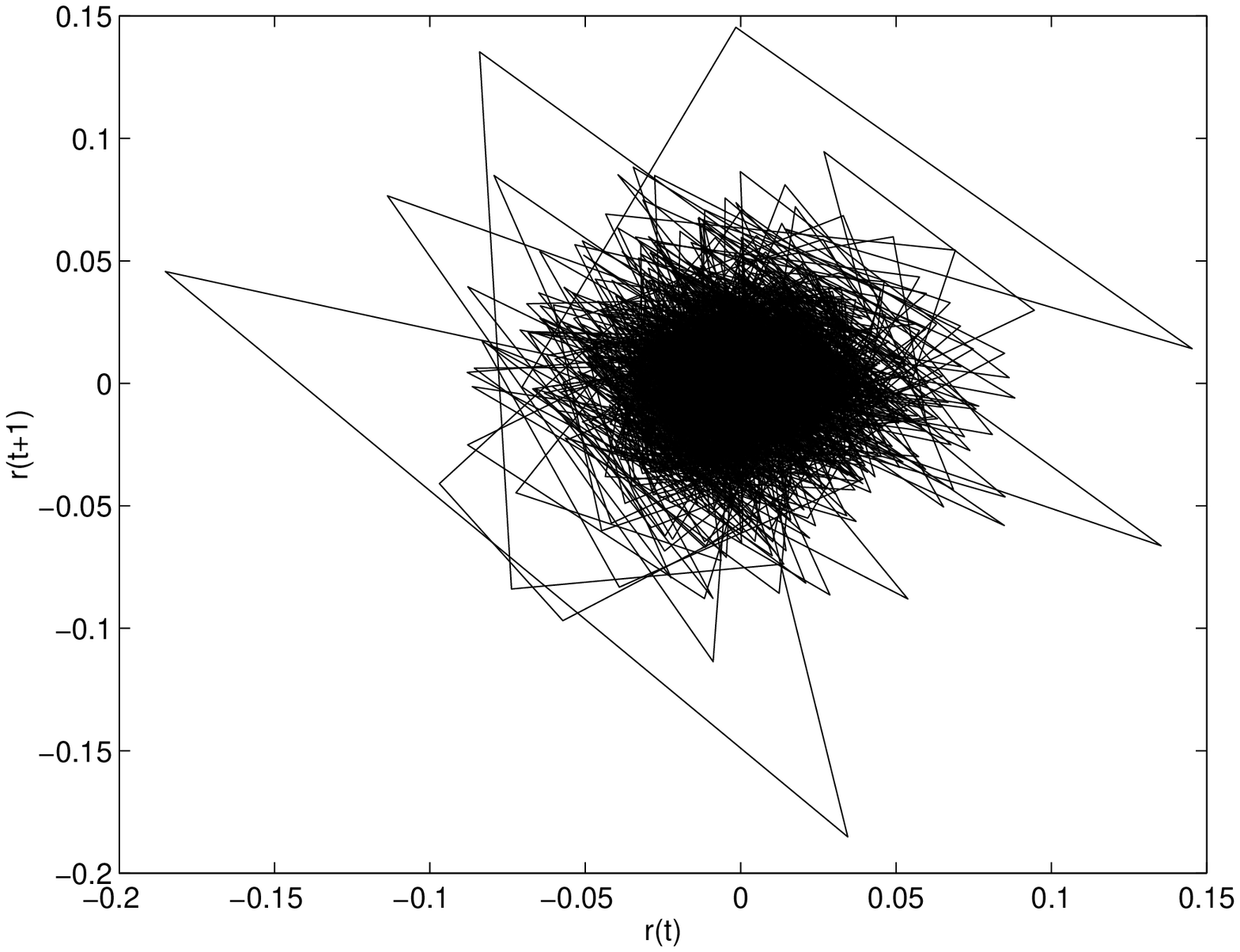,width=9truecm}
\end{center}
\caption{The dynamics of one-day returns}
\end{figure}

Fig.11 illustrates the dynamics of one-day returns 
\begin{equation}
r\left( t,1\right) \rightarrow r\left( t+1,1\right)  \label{3.7}
\end{equation}
It shows that the bulk of the data consists of a central core of small
fluctuations with a few large flights away from this core. This structure of
the data will have a strong influence on the results obtained in the next
section.

\section{Looking for a Gibbs measure}

Let us assume a coding of the dynamical system by a finite alphabet $\Sigma $%
. Then the space $\Omega $ of orbits of the system are infinite sequences $%
\omega =i_{1}i_{2}\cdots i_{k}\cdots $, $i_{k}\in \Sigma $, with the
dynamical law being a shift $\sigma $ on these symbol sequences. 
\begin{equation}
\sigma \omega =i_{2}\cdots i_{k}\cdots  \label{4.1}
\end{equation}
Depending on the dynamical law of the coded system, not all sequences will
be allowed. The set of allowed sequences in $\Omega $ defines the {\it %
grammar} of the shift. The set of all sequences which coincide on the first $%
n$ symbols is called a $n-${\it cylinder} (or $n-$block) and is denoted $%
[i_{1}i_{2}\cdots i_{n}]$. The probability measures over the cylinders is
the main tool that is used to characterize the dynamical properties and is a
piece of information that may be inferred from the data.

A particularly important measure on the cylinders is the Gibbs measure
defined by\cite{Bowen} \cite{Ruelle} 
\begin{equation}
c_{1}\leq \frac{\mu \left( [i_{1}(\omega )i_{2}(\omega )\cdots i_{n}(\omega
)]\right) }{\exp \left( -nP+\left( S_{n}\phi \right) (\omega )\right) }\leq
c_{2}  \label{4.2}
\end{equation}
with $\left( S_{n}\phi \right) (\omega )=\sum_{k=0}^{n-1}\phi \left( \sigma
^{k}\omega \right) $, $\phi $ being a H\"{o}lder continuous function on $%
\Omega $ called the {\it potential} and $P\left( \phi ,G\right) $ a function
depending on the potential and the grammar called the {\it pressure} of $%
\phi $.

The (equilibrium) Gibbs measure and the pressure bear an important
relationship to the entropy 
\begin{equation}
h\left( \mu \right) =\lim_{n\rightarrow \infty }\frac{H_{n}}{n}%
=\lim_{n\rightarrow \infty }\frac{1}{n}\sum_{i_{1}\cdots i_{n}}\mu \left(
[i_{1}i_{2}\cdots i_{n}]\right) \log \mu \left( [i_{1}i_{2}\cdots
i_{n}]\right)  \label{4.3}
\end{equation}
This is the variational principle that states that, for each potential and
grammar, the $\sup_{\eta }\left\{ h\left( \eta \right) +\int \phi d\eta
\right\} $ taken over all $\sigma -$invariant measures $\eta $ is reached
only for the Gibbs measure $\mu $ and equals the pressure 
\begin{equation}
P\left( \phi ,G\right) =h\left( \mu \right) +\int \phi d\mu  \label{4.4}
\end{equation}

The potential may be chosen in such a way that $P=0$. Such potential is
called a {\it normalized potential}. In this case we have the following
result\cite{Chazottes} 
\begin{equation}
\phi \left( \omega \right) =\lim_{n\rightarrow \infty }\log \frac{\mu \left(
[i_{1}(\omega )\cdots i_{n}(\omega )]\right) }{\mu \left( [i_{2}(\omega
)\cdots i_{n}(\omega )]\right) }  \label{4.4a}
\end{equation}
In principle this formula may be used to construct the potential using the
empirical measures $\widetilde{\mu }\left( [i_{1}(\omega )\cdots
i_{n}(\omega )]\right) $ obtained from the experimental sample. The problem
is that Eq.(\ref{4.4a}) requires the use of blocks of length $n$ as large as
possible but, for a finite sample, the statistics of such blocks suffers
from large uncertainties.

For practical purposes the most important class of Gibbs measures is the one
associated to finite range potentials, that is, functions on $\Omega $ that
depend only on the first $r$ symbols of a sequence $\omega \in \Omega $. The
importance of finite range potentials lies in the fact that they may be used
to uniformly approximate any H\"{o}lder continuous potential and, on the
other hand, given a limited amount of experimental data, only finite-range
potentials may be reliably inferred from experiment.

An important property of range$-r$ potentials is that for all values $%
i_{1}i_{2}\cdots i_{n}$ with $n\geq r$ \cite{Chazottes} 
\begin{equation}
\mu \left( [i_{1}\cdots i_{n}]\right) =\frac{\mu \left( [i_{1}\cdots
i_{r}]\right) \mu \left( [i_{2}\cdots i_{r+1}]\right) \times \cdots \times
\mu \left( [i_{n-r+1}\cdots i_{n}]\right) }{\mu \left( [i_{2}\cdots
i_{r}]\right) \mu \left( [i_{3}\cdots i_{r+1}]\right) \times \cdots \times
\mu \left( [i_{n-r+1}\cdots i_{n-1}]\right) }  \label{4.5}
\end{equation}
We will make use of this important relation in our attempt to look for a
Gibbs measure for the market fluctuation data. On the one hand the relation (%
\ref{4.5}) allows to express the entropy in terms of measures of cylinders
of finite length only, namely 
\begin{equation}
h\left( \mu \right) =-\sum_{i_{1}\cdots i_{k}}\mu \left( [i_{1}\cdots
i_{k}]\right) \log \frac{\mu \left( [i_{1}\cdots i_{k}]\right) }{\mu \left(
[i_{1}\cdots i_{k-1}]\right) }=H_{k}-H_{k-1}  \label{4.6}
\end{equation}
for all $k\geq r$ if $r>1$. If $r=1$ $h\left( \mu \right) =H_{1}$. $H_{k}$
is the entropy associated to cylinders of length $k$%
\begin{equation}
H_{k}=-\sum_{i_{1}\cdots i_{k}}\mu \left( [i_{1}\cdots i_{k}]\right) \log
\mu \left( [i_{1}\cdots i_{k}]\right)  \label{4.6a}
\end{equation}
This provides a criterium to find the range of the potential. Using the
empirical cylinder probabilities one computes $H_{k}$ for successively
larger $k$. Then, the range of the potential is found when $H_{k}-H_{k-1}$
tends to a constant value. Once the range is found, the potential may be
constructed directly from the empirical weights $\widetilde{\mu }\left(
[i_{1}\cdots i_{k}]\right) $.

Another important consequence of Eq.(\ref{4.5}) is that for $k>r$%
\begin{equation}
\mu \left( [i_{1}\cdots i_{k+1}]\right) =\frac{\mu \left( [i_{1}\cdots
i_{k}]\right) \mu \left( [i_{2}\cdots i_{k+1}]\right) }{\mu \left(
[i_{2}\cdots i_{k}]\right) }  \label{4.7}
\end{equation}
We will use both the criterium following from (\ref{4.6}) and Eq.(\ref{4.7})
to test for the possibility to construct a Gibbs measure for the market
fluctuation data.

A five-symbols code $\Sigma $%
\begin{equation}
\Sigma =\left\{ -2,-1,0,1,2\right\}  \label{4.8}
\end{equation}
is used for the one-day return data $r\left( t\right) $%
\begin{equation}
r\left( t\right) =\log p\left( t+1\right) -\log p\left( t\right)  \label{4.9}
\end{equation}
The average $\overline{r\left( t\right) }$ and standard deviation $s=\sqrt{%
\overline{\left( r^{2}\left( t\right) -\overline{r\left( t\right) }%
^{2}\right) }}$ of the returns are computed. Then, 
\begin{equation}
\begin{array}{clc}
\left( r\left( t\right) -\overline{r\left( t\right) }\right) >s & 
\Longleftrightarrow & 2 \\ 
s\geq \left( r\left( t\right) -\overline{r\left( t\right) }\right) >\frac{s}{%
3} & \Longleftrightarrow & 1 \\ 
\frac{s}{3}\geq \left( r\left( t\right) -\overline{r\left( t\right) }\right)
>-\frac{s}{3} & \Longleftrightarrow & 0 \\ 
-\frac{s}{3}\geq \sigma \left( r\left( t\right) -\overline{r\left( t\right) }%
\right) >-s & \Longleftrightarrow & -1 \\ 
-s\geq \left( r\left( t\right) -\overline{r\left( t\right) }\right) & 
\Longleftrightarrow & -2
\end{array}
\label{4.10}
\end{equation}

This coding is used and the empirical frequencies $\widetilde{\mu }\left(
[i_{1}\cdots i_{n}]\right) $ for blocks of successively larger order $n$ are
found. Of course $n$ cannot be arbitrarily large because of statistics.
Results will not be reliable whenever $5^{n}$ is larger than the size $N$ of
the data sample. The statistical reliability may be directly tested either
by comparing the number of different occurring blocks and $5^{n}$ or by
observing the fall-off of the empirically computed $H_{n}-H_{n-1}$.

\begin{figure}[htb]
\begin{center}
\psfig{figure=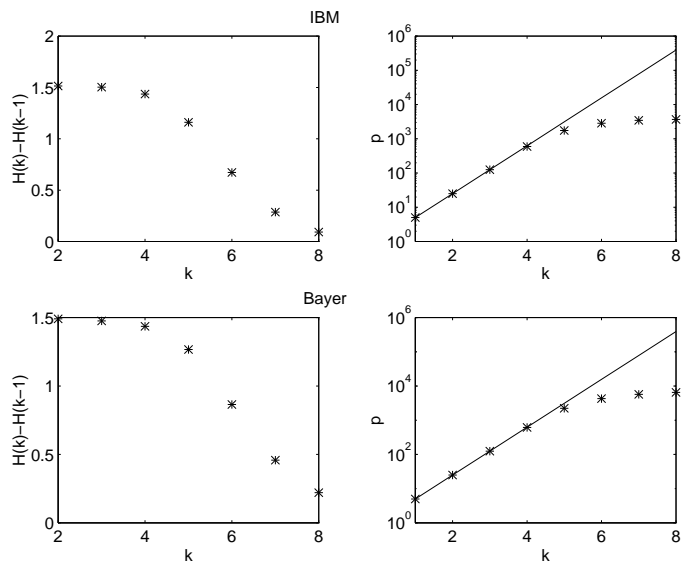,width=9truecm}
\end{center}
\caption{$H_{k}-H_{k-1}$ and the number of occurring blocks of size $k$ (IBM
and Bayer)}
\end{figure}

\begin{figure}[htb]
\begin{center}
\psfig{figure=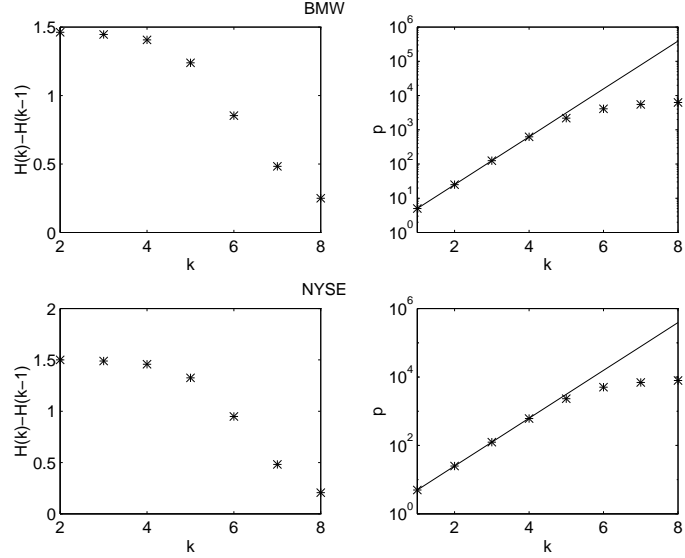,width=9truecm}
\end{center}
\caption{Same as Fig.12 for BMW and NYSE}
\end{figure}

First we try to estimate a possible range for the potential using the
criterium discussed above. The results are shown in Figs.12 and 13 for the
analyzed stocks and the NYSE index. The plots on the left show the quantity $%
H_{k}-H_{k-1}$ and the plots on the right compare the number $p\left(
k\right) $ of occurring blocks of size $k$ in the data with the maximum
possible number, $5^{k}$. Already for $k=2$ the difference $H_{k}-H_{k-1}$
seems to stabilize, staying nearly constant until $k=4$. After $k=4$ it
falls off, reflecting the lack of statistics also apparent in the comparison
of $p\left( k\right) $ with $5^{k}$ in the right hand side plots. These
results seem to suggest that the data is described by a very short range
potential. Notice that for a similar analysis performed on hydrodynamic
turbulence data the results are quite different with $H_{k}-H_{k-1}$ rising
smoothly up to a certain saturation level and then decreasing when one
reaches the lack of statistics level.

To check whether the short-range potential suggested by this criterium is
reliable or whether it simply results from some misleading feature of the
data, we have performed the test following from Eq.(\ref{4.7}). For
successively higher $k$ we estimate $\mu _{e}\left( [i_{1}\cdots
i_{k+1}]\right) =\frac{\widetilde{\mu }\left( [i_{1}\cdots i_{k}]\right) 
\widetilde{\mu }\left( [i_{2}\cdots i_{k+1}]\right) }{\widetilde{\mu }\left(
[i_{2}\cdots i_{k}]\right) }$ from the empirical $\widetilde{\mu }\left(
[i_{1}\cdots i_{k}]\right) $ and $\widetilde{\mu }\left( [i_{1}\cdots
i_{k-1}]\right) $, which is then compared with the empirically observed $%
\widetilde{\mu }\left( [i_{1}\cdots i_{k+1}]\right) $. The standard
deviation of the relative positive errors 
\begin{equation}
\varepsilon _{k}=\max \left( 0,\frac{\widetilde{\mu }\left( [i_{1}\cdots
i_{k+1}]\right) -\mu _{e}\left( [i_{1}\cdots i_{k+1}]\right) }{\frac{1}{2}%
\left( \widetilde{\mu }\left( [i_{1}\cdots i_{k+1}]\right) +\mu _{e}\left(
[i_{1}\cdots i_{k+1}]\right) \right) }\right)  \label{4.11}
\end{equation}
is computed and the number of blocks for which this error is one and two
standard deviations above the mean is computed. The result is plotted in
Fig.14 where the number of underestimation errors that are one (o) and two
(*) standard deviations away from the mean error are compared with the total
number $p(k)$ of different observed blocks of each length $k$.

\begin{figure}[htb]
\begin{center}
\psfig{figure=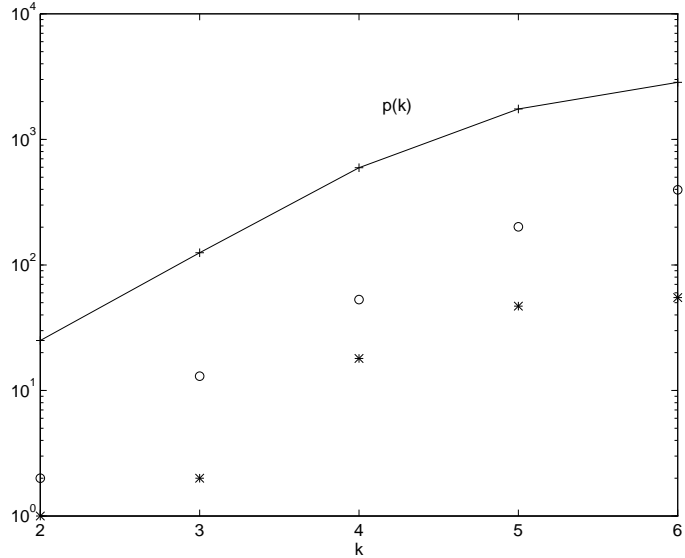,width=9truecm}
\end{center}
\caption{Underestimation errors one (o) and two (*) standard deviations away
from the mean and the total number $p(k)$ of observed blocks}
\end{figure}

One sees that the number of large deviation errors is very large and,
identifying the blocks for which these errors occur, one finds out that they
all correspond to blocks involving large positive or negative $r\left(
t\right) $'s ($2$ and $-2$). The conclusion is that a short-range potential
would describe the small fluctuations in the data, the large fluctuations
being badly described by it. The reason why the empirically found difference 
$H_{k}-H_{k-1}$ seems to saturate for a small $k$ is because, as is apparent
from Fig.11, the bulk of the data consists mostly of small fluctuations plus
a few large flights. The saturation of $H_{k}-H_{k-1}$ for small $k$ is a
reflection of the largely uncorrelated nature of the small fluctuations,
whereas other features like the large deviations, persistence of non-linear
correlations (volatility), etc. are not captured by a short-range potential.

Large deviations being misrepresented by an empirically constructed measure
is typical of situations where the actual measure is non-Gibbsian\cite
{vanEnter} \cite{Fernandez2}. In our case, however, it may also occur that
the measure is Gibbsian but with a long-range potential. This would
correspond to a sharp rise of $H_{k}-H_{k-1}$ at $k=2$ followed by a very
slow increase above $k=2$. In the empirical results a small increase may be
hidden by the fact that, as the block length increases, the statistics
becomes poorer. A large deviation analysis applied to the calculation of $%
H_{k}$ , using a standard technique\cite{Vilela1} to construct the free
energy and the deviation function from the data, is consistent with this
hypothesis.

In any case, whether a Gibbs measure exists or not, the finite-range
potential framework does not seem to be the more convenient way to describe
the market fluctuation process. In the next section we will explore another
approach specially suited to deal with long-memory processes.

\section{Market fluctuations as a chain with complete connections}

Processes with long memory have been studied in the past. Under certain
conditions, that is, when the dependence on the past does not decay too
slowly, existence and uniqueness of a well defined process may be proved. A
particularly well established framework is the one of chains with complete
connections and summable decays (\cite{Onicescu} \cite{Bressaud} \cite
{Comets} and references therein).

A stochastic process $\{X_{n}\}_{n\in {\Bbb Z}}$ with alphabet $\Sigma $ is
said to be a {\it chain with complete connections} (CCC) if the following
conditions are satisfied

\begin{enumerate}
\item  $\forall a_{i}\in \Sigma $%
\begin{equation}
P\left( X_{1}=a_{1},\cdots ,X_{n}=a_{n}\right) >0  \label{5.1}
\end{equation}

\item  The limit 
\begin{equation}
\lim_{m\rightarrow \infty }P\left( X_{0}=a_{0}|X_{j}=a_{j},-m\leq j\leq
-1\right) =P\left( X_{0}=a_{0}|X_{j}=a_{j},j\leq -1\right)   \label{5.2}
\end{equation}
exists $\forall a_{i},j\leq -1$

\item  There is a sequence $\left( \gamma _{m}\right) _{m\geq 1}$ with $%
\lim_{m\rightarrow \infty }\gamma _{m}=0$, such that for all $%
\{a_{j},b_{j}\in \Sigma ,j\leq -1\}$ with $a_{j}=b_{j}$ for $-m\leq j\leq -1$%
\begin{equation}
\left| \left( \frac{P\left( X_{0}=a_{0}|X_{j}=a_{j},j\leq -1\right) }{%
P\left( X_{0}=a_{0}|X_{j}=b_{j},j\leq -1\right) }-1\right) \right| \leq
\gamma _{m}  \label{5.3}
\end{equation}
\end{enumerate}

The process is said to be a {\it chain with complete connections and
summable decay} (CCCSD) if $\sum \gamma _{m}<\infty $

Conditions 1. and 2. are implicitly assumed when we considered the processes
(and pre-processed the data) to be asymptotically stationary. As for the
decays $\gamma _{m}$ they may be estimated from a typical sample of the
process. From the empirical probabilities for 
\begin{equation}
P\left( a_{0}|a_{1}\cdots a_{m}A\right) =\frac{P\left( a_{0}a_{1}\cdots
a_{m}A\right) }{P\left( a_{1}\cdots a_{m}A\right) }  \label{5.4}
\end{equation}
where $A$ is a block of arbitrary length, ones computes for each fixed set $%
a_{0}a_{1}\cdots a_{m}$ the maximum and the minimum over $A$, 
\begin{equation}
g\left( a_{0}a_{1}\cdots a_{m}\right) =\left( \frac{\max_{A}P\left(
a_{0}|a_{1}\cdots a_{m}A\right) }{\min_{A}P\left( a_{0}|a_{1}\cdots
a_{m}A\right) }-1\right)  \label{5.5}
\end{equation}
obtaining for $\gamma _{m}$%
\begin{equation}
\gamma _{m}=\max_{a_{0}a_{1}\cdots a_{m}}g\left( a_{0}a_{1}\cdots
a_{m}\right)  \label{5.6}
\end{equation}
However if the statistics for very long blocks is poor, which is in general
the case for finite samples, the computation of the maximum from empirical
data is not reliable. A better estimate of the decay behavior of the decay
rates is obtained from the following quantity, which smooths out the large
fluctuations due to poor statistics 
\begin{equation}
g(m)=\overline{g\left( a_{0}a_{1}\cdots a_{m}\right) }  \label{5.7}
\end{equation}
the average being taken over all sets $a_{0}a_{1}\cdots a_{m}$ of size $m$.

\begin{figure}[htb]
\begin{center}
\psfig{figure=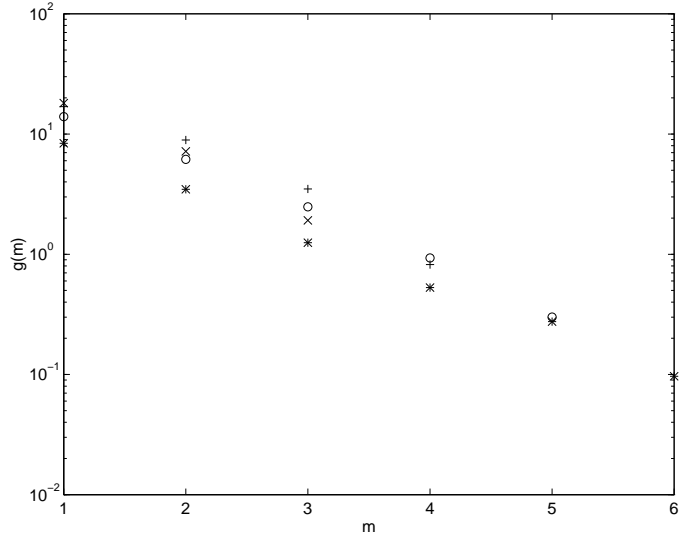,width=9truecm}
\end{center}
\caption{$g\left( m\right) $ computed using $A$ blocks of length 5 to 8 ($%
\times ,+,o,*$)}
\end{figure}

The results obtained for the $\Sigma -$coded data of the detrended
fluctuations (of BMW data) using $A$ blocks of length 5 to 8 ($\times ,+,o,*$%
) is plotted in Fig.15. Similar results are obtained for the other data. The
result is compatible with exponential decay, which would probably imply the
existence of a Gibbs measure (albeit with a long range potential). The data
for the maxima of $g\left( a_{0}a_{1}\cdots a_{m}\right) $ displays large
fluctuations and slower decay. However, with the amount of available data it
is not reliable for long blocks. In any case, in the present context of
CCC-processes, what the result suggests is the summability of the $\gamma
_{m}$'s. ($\sum_{m}\gamma _{m}<\infty $). For practical purposes the most
importance consequence of this fact is that a CCC-process with summable
decays is the $\overline{d}-$limit of its Markov approximations of order $k$%
. The nature of this approximation should however be clearly understood.

The $\overline{d}-${\it distance}\cite{Ornstein} between two processes
refers not to the processes themselves but to the process that implements
the coupling of the two processes. A {\it coupling} between two processes $%
X=\left\{ X_{n}\right\} $ and $Y=\left\{ Y_{n}\right\} $ over the alphabet $%
\Sigma $ is another process $\left\{ \widetilde{X}_{n},\widetilde{Y}%
_{n}\right\} $ defined over $\Sigma \times \Sigma $ such that the marginal
probabilities of $\widetilde{X}$ and $\widetilde{Y}$ coincide with those of $%
X$ and $Y$. Then the $\overline{d}-$distance between $X$ and $Y$ is 
\begin{equation}
\overline{d}\left( X,Y\right) =\inf \left\{ P\left( \widetilde{X}_{0}\neq 
\widetilde{Y}_{0}\right) :\left\{ \widetilde{X}_{n},\widetilde{Y}%
_{n}\right\} \textnormal{ is a stationary coupling of }X\textnormal{ and }Y\right\}
\label{5.8}
\end{equation}
For some types of coupling the two processes $\widetilde{X}$ and $\widetilde{%
Y}$ are know to coincide after a certain random time. However, for the
original processes $X$ and $Y$, if the $\overline{d}-$distance tends to zero
it does mean that the processes will coincide after a certain time. It only
means that it will occur for some other processes with the same marginal
probabilities.

This fact has an important bearing on the correct interpretation of the
``perfect simulation'' schemes \cite{Comets} proposed for CCC's. Perfect
simulation is always understood in the $\overline{d}-$distance sense and it
does mean perfect prediction. It means simply that a process is constructed
with the same conditional probabilities of the original process, whenever
the conditional probabilities of the original process are known. In practice
not all conditional probabilities involving infinite pasts are needed,
because going back to a regeneration time, only a finite number of back
steps are required. Several simulation schemes have been proposed for CCC's
with summable decays. The most important one for the applications, when the
conditional probabilities are inferred from experiment, is the sequence of
canonical Markov approximations of finite order $k$ ($k-$CMA). A $k-$CMA of
a process $X$ is a Markov chain $Y^{(k)}$ of order $k$ with conditional
probabilities $P^{(k)}$ such that 
\begin{equation}
P^{(k)}\left( a_{0}|a_{1}\cdots a_{k}\right) =P\left( a_{0}|a_{1}\cdots
a_{k}\right) =\sum_{A}P\left( a_{0}|a_{1}\cdots a_{k}A\right)  \label{5.9}
\end{equation}
For a CCC $X$ with summable decays\cite{Bressaud} 
\begin{equation}
\overline{d}\left( X,Y^{(k)}\right) \leq C\gamma _{k}  \label{5.10}
\end{equation}
$C>0$ being a constant. Actually the property of the Markov approximation
that is essential for the approximation result (\ref{5.10}) is 
\begin{equation}
\inf_{A}P\left( a_{0}|a_{1}\cdots a_{k}A\right) \leq P^{(k)}\left(
a_{0}|a_{1}\cdots a_{k}\right) \leq \sup_{A}P\left( a_{0}|a_{1}\cdots
a_{k}A\right)  \label{5.11}
\end{equation}
meaning that for Markov approximation schemes, other than the canonical one,
Eq.(\ref{5.10}) holds provided (\ref{5.11}) is satisfied. In fact, when the
conditional probabilities are inferred from limited experimental data a
different Markov approximation is more convenient.

The following approximation scheme is proposed for the market fluctuation
data, which we call the $\leq k-$Markov approximation:

i) Empirical transition probabilities $\widetilde{P}\left( a_{0}|a_{1}\cdots
a_{m}\right) $ are inferred from the occurrence probability of blocks of
order $m+1$. up to a certain order $m_{Max}$. Of course, only probabilities
that correspond to blocks $a_{1}\cdots a_{m}$ that appear in the data will
be available and especially for large $m$ many will be missing.

ii) For the simulation, with an approximation of order $k$, one looks at the
current block $\left( a_{1}\cdots a_{k}\right) $ of order $k$ and uses the $%
k-$empirical probability to infer the next state $a_{0}$. If that block has
not appeared in the data that was used to construct the empirical
probabilities, then one looks at the $k-1$ sized block $a_{2}\cdots a_{k}$
and uses the $k-1$ order empirical probabilities. If necessary the process
is repeated until an available empirical probability is found. This is the
reason why this is called the $\leq k-$Markov approximation.

This approximation scheme has been applied to the market fluctuation data
and for each $k-$order the successor $a_{0}$ of each block $\left(
a_{1}\cdots a_{k}\right) $ is compared with a prediction $\widetilde{a}_{0}$
obtained by throwing a random number with the probabilities $\widetilde{P}%
\left( a_{0}|a_{1}\cdots a_{k}\right) $. Figs.16 shows some of the results.
In all cases the quantity that is plotted is the averaged squared error 
\begin{equation}
e^{2}=\left\langle \left( \widetilde{a}_{0}-a_{0}\right) ^{2}\right\rangle
\label{5.12}
\end{equation}
the average being taken over the samples and 100 different runs. The two
upper plots and the left lower plot show the results obtained (for each
approximation order $k$) when half of the data for each company is used to
predict the other half. The points labelled ($o$) correspond to the past
used to predict the future and those labelled ($*$) to the future used to
predict the past. Finally the right lower plot shows the results obtained
when $\widetilde{a}_{0}$ is chosen at random (for the 3 companies, IBM $%
\left( o\right) $, Bayer $\left( *\right) $ and BMW $\left( +\right) $). The
main conclusions that may be extracted from these results are:

\begin{figure}[htb]
\begin{center}
\psfig{figure=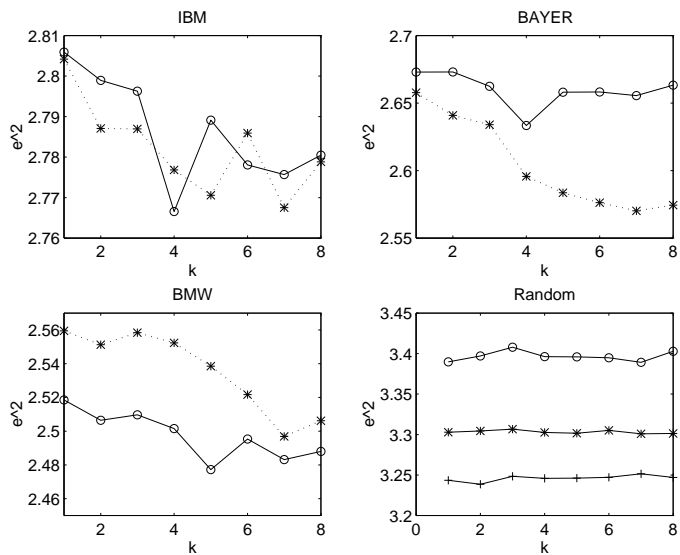,width=9truecm}
\end{center}
\caption{The past predicting the future ($o$) and the future predicting the
past ($*$), compared to random choice}
\end{figure}

\begin{itemize}
\item  The average prediction obtained from using the empirical
probabilities is better than a random choice.

\item  However, the main improvement is a result of a correct accounting of
the two-symbol probabilities ($k=1$).

\item  After the improvement due to the use of the lowest order blocks a
small (but consistent) improvement is found by using the past information up 
$k=4$ or $5$. No significant improvement is obtained by using higher order
approximations. This is consistent with the poorer statistics of large
blocks. Actually for each individual simulation the result of using $k>5$
leads to much larger fluctuations.\medskip 
\end{itemize}

The main conclusion is that although the bulk of the data is represented by
a short-memory process, there is nevertheless evidence for a small
long-memory component that is captured by the higher-order Markov
approximations. Depending on the amount of data that is available to infer
the empirical conditional probabilities there is a maximum $k=k_{m}$ that
should be used for the simulation process. This $k_{m}$ value may be
estimated from the quantity $p\left( k\right) $ plotted in Figs 12 and 13.
Finally, although a mild gain is obtained from using $k_{m}-$block
probabilities rather than one-symbol probabilities, it should be remembered
that perfect simulation in the $\overline{d}-$distance sense is not perfect
prediction for the actual process. This is a point to keep in mind when
attempting to develop any trading strategies based on the empirical block
probabilities.

We have also explored the use of the empirical probabilities of one company
to predict the behavior of the others. In all cases the improvement coming
from the one-symbol probabilities (as compared to random choice) is
obtained. This means that the one-symbol probabilities are similar in all
companies. However for the long-memory component the behavior is very much
company-dependent. For example there seems to be no correlation of this
component between IBM and the other two companies, with the prediction being
actually worse when the empirical probabilities for longer blocks are used.
The same happens also when the empirical probabilities of BMW and Bayer are
used to predict IBM. However there is some statistical correlation between
the long-memory components (and some mild prediction improvement) between
BMW and Bayer. This suggests that the statistical short-memory component of
the market process might be similar for many different stocks, whether the
long-memory component might be different from market to market and to divide
the stocks into classes. A similar conclusion follows from the stocks
taxonomy obtained by Mantegna \cite{Mantegna1}, although that work does not
distinguish between the short- and long-memory components of the process.

\section{Conclusions}

\begin{enumerate}
\item  The bulk of the market fluctuation process seems to be a short-memory
process. In addition it has a small long-memory component, which however is
very important for practical purposes because it is associated with the
large fluctuations of the returns.

\item  Existence of the long-memory component suggests the {\it chains with
complete connections and summable decays} as the appropriate framework to
describe these processes. Although the decays may be exponentially
converging, the lack of accurate data concerning long blocks prevent an
accurate description by a finite range Gibbs potential.

\item  The sequence of empirical based $\leq k-$Markov approximations
discussed in Sect.5 seems the most unbiased simulation of the process.
Eventual convergence in the $\overline{d}-$distance sense is expected to
hold because the market fluctuation process seems to fit in the framework of
chains with complete connections and summable decays.

\item  Except for cases where one is sure of the existence of a finite
potential, Markov approximations must always be used if only finite data is
available. This true whether a Gibbs measure exists or not. What the chains
with complete connections framework provides though, is a rationale for the
convergence of the Markov approximations and a criterium to estimate,
through the $\gamma _{m}$ decays, how good this approximation is. Notice
however the trade-off between higher order approximations and lack of
statistics, that leads to an optimal block length for the empirical
probabilities to be used in the simulations.

\item  As work for the future we point out that it would be interesting to
analyze in this framework the high frequency market data. Here however
attention should be paid to the possibly multi-scale and multi-component
nature of the processes.
\end{enumerate}

\end{document}